\documentclass[journal]{vgtc}                     %

\newcommand{\add}[1]{\textcolor{black}{#1}}

\onlineid{1258}

\vgtccategory{Research}

\vgtcpapertype{Analytics \& Decisions}

\title{
Beyond Correlation: Incorporating Counterfactual Guidance to \\ Better Support Exploratory Visual Analysis
}

\author{
  \authororcid{Arran Zeyu Wang}{0000-0002-7491-7570}, \authororcid{David Borland}{0000-0002-0162-4080}, and \authororcid{David Gotz}{0000-0002-6424-7374}
}

\authorfooter{
  \item Arran Zeyu Wang and David Gotz are with the University of North Carolina at Chapel Hill. 
  E-mail: zeyuwang@cs.unc.edu, gotz@unc.edu
  \item David Borland is with RENCI at the University of North Carolina at Chapel Hill.
  E-mail: borland@renci.org
}

\shortauthortitle{Counterfactual Guidance in Exploratory Visual Analytics}

\teaser{
\centering
\vspace{-2mm}
\includegraphics[width=0.995\columnwidth]{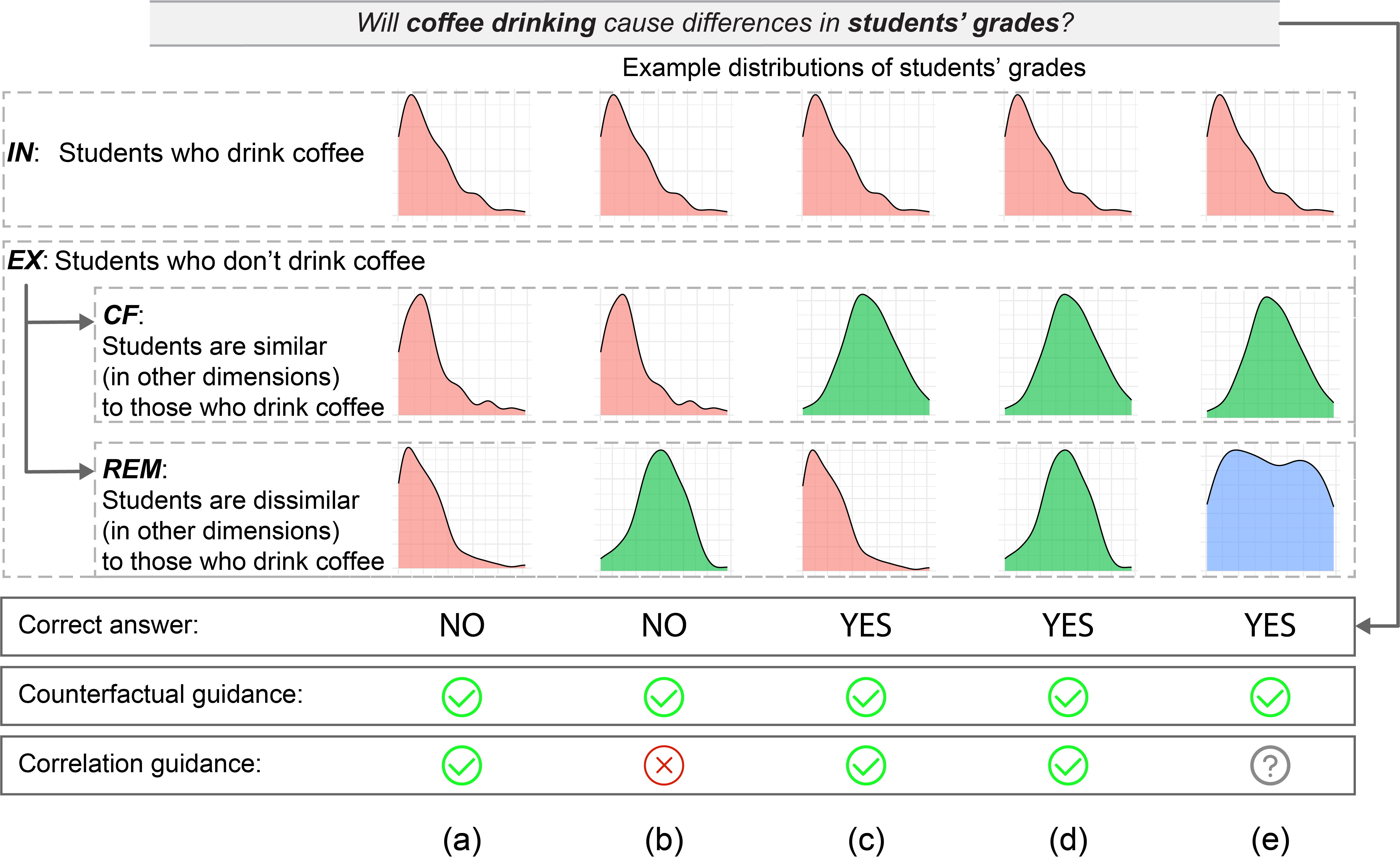}
\vspace{-7mm}
\caption{
The proposed counterfactual guidance technique can be compared with traditional correlation-based guidance by considering five archetypal scenarios.
Consider the example question ``\textit{Will coffee drinking cause differences in students' grades?}'' An analyst might compare data based on whether students drink coffee or not, and attempt to answer the question based on differences in the distribution of grades for the resulting subsets.
The leftmost column lists the subsets created in this process (see \autoref{sec:definesubset} for details), and the 
various charts illustrate five potential combinations (a-e) of distributions across the different subsets which suggest different possible answers to the analytical question (see \autoref{sec:archetypes} for details). 
Across the bottom of the figure, the symbols indicate which methods more accurately reflect the correct interpretation of the data. As the example illustrates, counterfactual-based approaches have advantages in two of the five scenarios while they perform equally on the other three.
}
  \vspace{-2mm}
\label{fig:teaser}
}

\abstract{
Providing effective guidance for users has long been an important and challenging task for efficient exploratory visual analytics, especially when selecting variables for visualization in high-dimensional datasets. Correlation is the most widely applied metric for guidance in statistical and analytical tools, however a reliance on correlation may lead users towards false positives when interpreting causal relations in the data. In this work, inspired by prior insights on the benefits of counterfactual visualization in supporting visual causal inference, we propose a novel, simple, and efficient counterfactual guidance method to enhance causal inference performance in guided exploratory analytics based on insights and concerns gathered from expert interviews. Our technique aims to capitalize on the benefits of counterfactual approaches while reducing their complexity for users. We integrated counterfactual guidance into an exploratory visual analytics system, and using a synthetically generated ground-truth causal dataset, conducted a comparative user study and evaluated to what extent counterfactual guidance can help lead users to more precise visual causal inferences. The results suggest that counterfactual guidance improved visual causal inference performance, and also led to different exploratory behaviors compared to correlation-based guidance. Based on these findings, we offer future directions and challenges for incorporating counterfactual guidance to better support exploratory visual analytics.
\vspace{-0.5mm}
} %

\keywords{Counterfactual, Guidance, Exploratory visual analysis, Visual causal inference, Correlation
\vspace{-3mm}}

\graphicspath{{figs/}{figures/}{pictures/}{images/}{./}} %

\usepackage{tabu}                      %
\usepackage{booktabs}                  %
\usepackage{lipsum}                    %
\usepackage{mwe}                       %

\usepackage{mathptmx}                  %
\usepackage{amsmath,amsfonts}
\usepackage{algorithmic}
\usepackage{algorithm}
\usepackage{array}
\usepackage[caption=false,font=normalsize,labelfont=sf,textfont=sf]{subfig}
\usepackage{textcomp}
\usepackage{stfloats}
\usepackage{url}
\usepackage{verbatim}
\usepackage{graphicx}
\usepackage{cite}  
\usepackage{amsmath}

\begin{document}

\maketitle
\section{Introduction}
\label{sec-intro}

Supporting efficient discoveries of insights within complex datasets is a primary goal for exploratory data analysis.
Visual analytics tools often employ guided approaches to lead users to find meaningful inferences from high-dimensional data~\cite{tukey1977exploratory, ceneda2019review, gotz2008characterizing}.
The most typical and widely applied guidance metric is correlation~\cite{wilkinson2010systat, Batt2020Learning}, 
however correlation-based guidance may mislead users by suggesting false causal relationships~\cite{tukey1977exploratory, pearl2009causality, borland2024using}.

Filtering is a common step in visual analytics workflows~\cite{keim2008visual, card1999readings, shneiderman1996eyes}, whereby users can create data subsets of interest based on specified constraints to help answer analytical questions. However, ad-hoc filtering operations can also lead to mistaken assumptions regarding the strength and causal nature of relationships between variables. 
Recent advances have employed \emph{counterfactuals} in visualization---visualizing additional data subsets designed to provide improved context---to provide benefits to various visualization and visual analytics tasks, such as better interpretations of machine learning models~\cite{wexler2019if, gomez2020vice, cheng2020dece} and improved visual causal inference~\cite{kaul2021improving, wang2024empirical, wang2024framework}.
However, counterfactual-based methods require more complicated and nuanced interpretation of visualizations, potentially leading to more time-consuming and complex analyses~\cite{borland2024using}.

This paper aims to capitalize on the benefits of counterfactual approaches, while also reducing complexity for the user, by using counterfactuals to improve guidance in visual analytics systems with respect to causal interpretations of data.
Inspired by existing insights and expert interviews, we introduce a novel counterfactual-based guidance technique designed to capture differences between subsets created by counterfactual visualization techniques~\cite{wang2024framework}.
Similar to correlation-based methods, our approach outputs a numeric value that can be used to guide users' exploration, thus simplifying the complexity typically associated with explaining counterfactuals to users.
Our technique therefore combines the benefits of counterfactuals to better support guided visual exploration while mitigating its limitations.
\add{Unlike previous counterfactual visualization work, our study incorporates counterfactuals into a guidance technique that enables effective exploration of datasets while significantly reducing visualization complexity. In addition, we provide a more thorough analysis of users' exploratory patterns.}

We illustrate the benefits of counterfactual guidance compared to correlation-based guidance via a theoretical scenario using a simple students' coffee drinking example (\autoref{fig:teaser}). Through this use case, we show how counterfactual guidance can avoid incorrect inferences and more effectively lead to correct inferences compared to correlation-based guidance.

Furthermore, through a comparative user study with a prototype exploratory visual analysis system using a synthetic dataset with ground-truth causal relationships, we demonstrate that counterfactual guidance leads to improved performance in visual causal inference tasks compared to correlation-based guidance.
Based on the findings, we propose design implications for better integrating counterfactual guidance into exploratory visual analytics systems, aiming to facilitate more efficient and insightful data exploration.

Specifically, the contributions of this paper include:
\begin{itemize}
    \item {\bf A counterfactual-based guidance technique with an open-sourced library to support exploratory visual analysis.} We propose a new counterfactual-based method to compute guidance for visual analytics systems. Furthermore, we provide an open-source Python library to compute counterfactual guidance.

    \item {\bf Theoretical and empirical evidence demonstrating the benefits of counterfactual guidance for visual analysis.} We demonstrate a theoretical use case and present results from an empirical user study to illustrate the benefits of counterfactual guidance versus traditional correlation-based guidance.
    
    \item {\bf Reflecting on prior work and discussing future research directions.} We discuss how our study can reflect and confirm prior insights and indicate future research directions to better incorporate counterfactual guidance with exploratory analytics.

\end{itemize}

\section{Background and Related Work}
\label{sec-related}

The methods presented in this paper build on prior work in two broad areas of related research. First, our approach is informed by prior work exploring counterfactuals and their applications in support of visual causal inference.  Second, our contributions are designed to extend previous approaches to guided exploratory visual analysis. 

\subsection{Counterfactuals in Visual Causal Inference}

\add{Causal inference techniques are designed to help characterize the causal relationships between various factors within a dataset, showing how one factor may lead to changes in another.}
Pearl~\cite{pearl2009causality} established \emph{counterfactual reasoning} as the most advanced level of his proposed statistical causal inference hierarchy, involving the exploration of hypothetical alternatives to observed events.
\add{The statistical and machine learning communities have proposed many techniques to support causal inference.
For example, instrumental variables have been employed to explore causal structures among datasets~\cite{angrist1996identification}.
In other work, machine learning approaches were utilized to perform causal inference from complex data~\cite{johansson2016learning}.
Alternatively, score matching methods can be used to extract impact factors for target outcomes to help create causal models~\cite{stuart2010matching, iacus2012causal}.}

Due to their utility in exploring outcome relations~\cite{morgan2015counterfactuals}, counterfactuals have been increasingly applied in visualization research to enhance the understanding of datasets~\cite{borland2024using}.
\add{For example, Kaul et al. introduced the first general-purpose counterfactual-based exploratory visual analytics system, \emph{CoFact}~\cite{kaul2021improving}.
Through a user study, they found that \emph{CoFact} could assist users in inferring feature-to-outcome relations from datasets.
However, \emph{CoFact} utilized complex counterfactual visualizations without well-designed guidance techniques, making effective data exploration challenging.}

Further, preliminary research has explored the potential of counterfactual visualization---visualizing data subsets that do not match filter inclusion criteria, but are similar to the included subset in other ways---to benefit users' causal inferences in exploratory tasks.
\add{Wang et al. proposed a causality comprehension model and found that counterfactual visualizations benefit users' causality comprehension for juxtaposed visualizations~\cite{wang2024empirical}.
However, their studies focused on static statistical charts and did not address the use of counterfactuals in an exploratory context.}

However, challenges remain in effectively conveying these complex causal relationships through visual means~\cite{borland2024using}. The most effective ways to present information in order to improve causal inference is still an area of active research.
Prior studies also found that although visualizing counterfactuals can provide improved performance in causal inference, due to their increased visual and conceptual complexity users typically took longer to explore and interpret them~\cite{wang2024empirical}.
These studies are also limited by a lack of ground truth causal relationships to validate any advantages of counterfactuals.

These approaches have shown promise in helping users form more accurate interpretations of data, although the field is still exploring the most effective ways to integrate counterfactuals into visual data communication.
Built upon existing insights, we aim to maintain the benefits of visualizing counterfactuals while mitigating their increased complexity in interpreting visualizations.
In this paper, we present a simple yet effective counterfactual measure to guide exploratory visual analysis.

\subsection{Guided Exploratory Visual Analysis}

Guided exploratory analysis in visual analytics refers to the process of leading users through a structured exploration of data to uncover insights~\cite{ceneda2019review}. 
Correlation between data variables
is the most widely applied metric to guide users towards potentially interesting insights for exploratory visual analysis, \add{and has been applied in various domains such as statistical software~\cite{wilkinson2010systat}, visual analysis tools~\cite{Malik2012A}, and biostatistics methodologies~\cite{chan2003biostatistics}.}

Ceneda et al. explored a taxonomy of guidance in the context of visual analytics~\cite{ceneda2016characterizing}.
They emphasized that the major goal of such guidance is to mitigate the effects of the knowledge gap across different guidance degrees~\cite{ceneda2018guidance}.
They further developed theoretical frameworks to better characterize guidance in visual analytics through analyzing designers' requirements~\cite{ceneda2020guide}, descriptively connecting visualization onboarding and guidance~\cite{stoiber2022perspectives}, and specifying practical guidance strategies~\cite{sperrle2022lotse}.

Other approaches have explored practical methods for guided visual analysis.
The progressive visual analytics workflow~\cite{stolper2014progressive} enables user exploration of partial results to quickly lead to the next exploration step by inferring early and meaningful clues.
Feedback-driven visual analytics~\cite{behrisch2014feedback} can provide benefits by providing relevant feedback in guiding users during the analysis of large multidimensional datasets.
\emph{SOMFlow}~\cite{sacha2017somflow} enables guided exploration for cluster analysis using time-series self-organizing maps.
\emph{EVM}~\cite{kale2023evm} incorporates model checks~\cite{hullman2021designing} into visual analytics systems to guide users in better examining the efficiency of data exploration and interpretation based on statistical models.
Indexing~\cite{guo2023does} and faceted~\cite{guo2023grafs} guidance approaches were reported to improve users' exploration efficiency in exploratory search tasks.
AI-supported guidance~\cite{ha2024guided} can also benefit users' trust and exploration during visual analysis, especially in more difficult tasks.

Recommendation techniques have also been shown to be beneficial for visual analytics guidance~\cite{zhou2022design}.
For example, modeling user behavior~\cite{gotz2009behavior} and analytical focus during visual analysis~\cite{zhou2021modeling} can lead to improved user exploration in various usage scenarios such as mass text document analysis~\cite{gotz2010harvest}, web search~\cite{cheng2009context}, data pre-fetching~\cite{battle2016dynamic}, and combating bias~\cite{gotz2016adaptive}.
Task-driven approaches for recommendations have also been employed to guide advanced mixed-initiative visual analytics of users~\cite{cook2015mixed}.
Other recommendation systems~\cite{wongsuphasawat2016voyager, moritz2018formalizing, wang2024automated} utilize design principle-driven recommendations to efficiently guide user exploration in exploratory visual analytics.

Although this breadth of research offers significant insights on how to effectively guide user exploration for various specific scenarios, correlation-based guidance is still the most widely applied in analytical and statistical software such as SYSTAT~\cite{wilkinson2010systat} and Tableau~\cite{Batt2020Learning}.
To our knowledge, there have been no prior studies examining the use of counterfactual guidance for visual analytics systems.
In this work, we compare counterfactual guidance to correlation-based guidance with respect to their performance supporting causal inference.

\section{Archetypal Usage Scenarios}
\label{sec:usage}

In this section, we describe five archetypal scenarios for counterfactual and correlation-based guidance in data analysis. Through these scenarios, we aim to demonstrate how and when counterfactual guidance can offer benefits over correlation-based methods.

\subsection{Data Subsets}
\label{sec:definesubset}

First we briefly introduce the definitions of data subsets related to computing counterfactuals 
to enable counterfactual guidance. For more detailed definitions see \cite{wang2024empirical}.

\paragraph{\textbf{IN:}} The included (IN) subset comprises the data samples that match user-chosen inclusion criteria when filtering a dataset.

\paragraph{\textbf{EX:}} The excluded (EX) subset comprises the data samples that do not match the inclusion criteria for IN.

\paragraph{\textbf{CF:}} The counterfactual (CF) subset comprises data samples from EX that are the most similar to those from IN based on variables in the data other than the inclusion criteria.
Following prior studies~\cite{wang2024empirical, kaul2021improving}, we employ the Euclidean distance as the default similarity measure.  

\paragraph{\textbf{REM:}} The remainder (REM) subset comprises the data samples from EX that are not included in CF.

\autoref{fig:teaser} illustrates instances for these subsets.
In this case, the corresponding data subsets refer to: 
\textbf{IN:} Students who drink coffee.
\textbf{EX:} Students who don't drink coffee.
\textbf{CF:} Students who don't drink coffee but are similar to IN across other variables.
\textbf{REM:} Students who neither drink coffee nor are similar to IN across other variables.

In the following sections, we employ $D_{IN, CF}$ and $D_{IN, REM}$ as terms to represent the differences between IN and CF, and between IN and REM, respectively. We also refer to a low guidance value as one that reflects a low evidence for a causal effect of the filter variable on the outcome, and a high value as one that reflects greater evidence for a causal effect.

\subsection{Archetypes Overview}
\label{sec:archetypes}

The five archetypes presented in this section are determined based on the degree of similarity between the key subsets defined earlier in this paper: IN, CF, and REM. More specifically, if we take a simplified binary view of similarity between two subsets, we can specify five different scenarios relating these three sets: (1) IN is similar to both CF and REM; (2) IN and CF are similar, but REM is different; (3) IN and REM are similar, but CF is different; (4) CF and REM are similar, with both different from IN; and (5) all subsets are different from each other.

We illustrate these five archetypal scenarios with examples from a guided exploratory analysis of data describing how coffee drinking relates to students' grades as depicted in \autoref{fig:teaser}. In all of these examples, the filter \textit{students who drink coffee} is used to define IN while the distributions represent the outcome variable \textit{students' grades}.

\subsection{Case 1: All Subsets have Similar Distributions}
When all subsets exhibit similar data distributions (\autoref{fig:teaser} (a)), the conclusion is relatively clear: there is no evidence that the filter variable (coffee consumption) meaningfully influences the outcome (students' grades). This can be seen by comparing the grades of those who consume coffee with those who don't (both similar and non-similar students). In all cases, the grades are similarly distributed.

Mathematically, this scenario will lead to a very low $D_{IN, CF}$ value as well as a very low $D_{IN, REM}$ value, and should result in a low counterfactual guidance value.
Similarly, the correlation between coffee drinking and grades (as evidenced by the small difference in outcomes between IN and EX) is low, resulting in a correspondingly low correlation guidance value.

\subsection{Case 2: REM is Different} In this case, IN and CF exhibit similar outcomes (similar grades), while the REM subset shows a different outcome distribution (\autoref{fig:teaser} (b)).  This case suggests that 
students who drink coffee earn grades similar to those of non-coffee drinkers who are "just like them" except for their coffee drinking. In contrast, students in the REM subset 
do not drink coffee but are also dissimilar from coffee-drinking students in other ways beyond their coffee intake.  The REM students, in this case, earn different grades from the others and these can be attributed to factors other than coffee given the similar grades within IN and CF.

For these reasons, a user would ideally avoid focusing on these types of variables, and guidance during exploratory analysis would not push users toward such a pattern.
When considering counterfactual guidance approaches, 
even though $D_{IN, REM}$ can be relatively high, guidance should still be low because of the low $D_{IN, CF}$ value. In sharp contrast, 
correlation-based guidance might very well lead users directly to this less interesting pattern because the correlation between coffee drinking and grades (as evidenced by the difference between IN and EX) is substantial given that $REM$ is part of $EX$ along with the large $D_{IN, REM}$. 

\subsection{Case 3: CF is Different}
A third possible pattern shows IN and REM having similar outcomes while CF is different as shown (\autoref{fig:teaser} (c)). This case is less common and reflects a more complex circumstance.  Using the coffee and grades example, the difference in students' grades between IN and CF suggests that coffee drinking is an important factor given that IN and CF contain similar students except for their coffee consumption.  However, the similar grades between IN and REM, which also differ in coffee consumption, suggest that other factors that distinguish between REM and CF may also influence the students' grades. Moreover, the other factors may influence grades in a way that is similar to coffee consumption.

This pattern could be reflected in both counterfactual and correlation-based guidance approaches.  Counterfactual guidance would lead users to explore these cases because of the large $D_{IN, CF}$ value. Correlation guidance, meanwhile, would capture the difference between IN and EX, though the strength of the signal may be weaker due to it reflecting a combination of the large $D_{IN, CF}$ and the low $D_{IN, REM}$.

\subsection{Case 4: IN is Different} A fourth common pattern is when the CF and REM subsets have similar outcomes that are both different from the IN subset (\autoref{fig:teaser}~(d)). In our example, this pattern would reflect coffee drinkers earning grades that are different from those who don't drink coffee and that this difference was seen for all non-coffee drinkers regardless of how similar or different the students were to their coffee-drinking counterparts. 

In this case, the difference in students' grades for both REM and CF to IN will result in large values for both $D_{IN, REM}$ and $D_{IN, CF}$.
Therefore, this pattern would be identified by both correlation-based guidance and counterfactual guidance. 

\subsection{Case 5: All Subsets are Different}
In the fifth and final case, the outcome distributions are different across all three subsets IN, CF, and REM (\autoref{fig:teaser}~(e)). This situation would reflect that all three groups have different grade distributions: coffee-drinking students, non-coffee-drinking students who are like their coffee-drinking peers, and non-coffee-drinking students who are dissimilar from their coffee-drinking peers.

This scenario combines aspects from both cases~2 and~3. As in case~3, counterfactual guidance would highlight this pattern for exploration given the large difference between IN and CF. However, as in case~2, CF and REM are also different. The correlation-based guidance would therefore be more difficult to predict because it depends on the combined distributions of those two subsets. 

\subsection{Summary}

As outlined in the description of these five archetypes and their depiction in \autoref{fig:teaser}, correlation-based and counterfactual guidance can both be used effectively under multiple conditions. More specifically, they can both be used to help productively guide users towards cases~3 and~4 and away from case~1. In contrast, cases~2 and 5 are more problematic for correlation-based guidance and could result in incorrect or misleading guidance, whereas counterfactual guidance should be able to provide correct results for both cases.

These archetypes help demonstrate the theoretical rationale for, and benefits of, counterfactual guidance as summarized in the table at the bottom of \autoref{fig:teaser}. Motivated by these observations, the study presented in \autoref{sec-evaluation} provides empirical evidence about the benefits of counterfactual guidance during exploratory analysis when compared to a correlation-based approach.

\section{Counterfactual Guidance}
\label{sec-counterfactual}

In this section, we first share results from formative interviews with visualization experts which aimed to distill key design requirements. Then, informed by the findings from those interviews we present the details of our counterfactual guidance methodology.

\subsection{Insights and Concerns from Expert Interviews}
\label{sec-design}

Building upon prior findings on counterfactuals in visualization, we gathered suggestions from qualitative expert interviews to identify rationales and practical key insights
to incorporate counterfactuals into visual analysis systems.
The interviews were conducted with 6 experts, comprising three visualization researchers and three visualization engineers. The interviews lasted 30 minutes on average.

During each interview, experts were introduced to and shown existing counterfactual visualizations and visual analytic tools. They then discussed their concerns and proposed suggestions for building efficient and easy-to-use counterfactual-based analytics systems.
We summarize the main insights and concerns raised during these interviews within the following themes.

\paragraph{Complex counterfactual concepts.}
One common theme was that counterfactual visualizations were seen as helpful tools, but that they are not typically employed in their analytical workflows and may be a complex concept for new users to understand.
Four out of six experts reported they had no prior knowledge of counterfactuals, and that understanding the definitions involved in counterfactual visualizations was challenging for them even though examples were provided.
All experts suggested that if we want to effectively incorporate counterfactuals into the visual analytics workflow for general users, it would be best if it was done in a way that avoids introducing new complex concept definitions.

\paragraph{Difficulty in understanding the impact of the REM subset.}
Another challenge raised by the experts focused on the REM subset.
When exploring datasets, users are primarily focused on the impact of chosen filters, i.e., the IN subset.  Moreover, the experts all agreed that visualizing the CF subset was useful as a way to help them understand the differential impact of their chosen filters.
However, five experts mentioned that they didn't easily understand the usefulness of looking at data from the REM subset.  This aligns with the more complex conceptual basis required to meaningfully account for the REM subset when interpreting a chart. It involves a three-way subset comparison which requires a deeper understanding of counterfactuals (e.g., 'if IN and CF are different, but REM and CF are alike...').
Given this difficulty, five experts suggested that visualizing REM be de-emphasized.  This reflects a difficulty in implementation, however, as the REM subset is critical in counterfactual interpretation.

\paragraph{Lack of simple explanatory use cases.}
Previous studies provided use cases to explain how counterfactuals in visualizations could be interpreted. However, five experts expressed concern that these use cases may be too complex for users to understand.
Furthermore, they reported that the examples didn't effectively illustrate at a glance how counterfactuals can be useful for data exploration rather than the interpretation of an individual chart. This concern helped guide the development of the archetypes in \autoref{sec:usage}.

\paragraph{Complexity in visualizations.}

Many current counterfactual visualization systems typically combine multiple charts to render the different subsets.
However, four experts suggested that this kind of presentation significantly increases the complexity of interpreting visualizations, especially for general users. Users have to interpret the individual charts and then mentally combine them to draw the correct insight based on their understanding of counterfactual reasoning.
Therefore, our experts suggested that we find ways to simplify the representations of counterfactual visualizations.

\subsection{Computing Counterfactual Guidance}

Given these insights, we aimed to use counterfactual information in a manner that hides some of the complexity from users. By calculating a value based on user-selected filters in the computed data subsets, counterfactual information can be used to provide improved guidance for exploratory visual analytics. Examples illustrating the counterfactual guidance metric are provided in \autoref{sec:usage}.

Note that all the guidance computations described here incorporate only user-selected variables (filters and outcome) and do not include other dataset variables that are used by the computation of data subsets (see \autoref{sec:definesubset}).

\subsubsection{Counterfactual Dissimilarity}

The counterfactual guidance is based on the similarity under user-selected variables for filters and the outcome of interest (e.g., \emph{coffee drinking} and \emph{students' grades} in \autoref{fig:teaser}) between the previously introduced data subsets.
To compute the distance between data subsets, we first define the $Similarity$ between two individual data points as:
\begin{equation}
\label{eq:dissimilarity}
    Similarity(i,j) = exp^{-distance(i,j)} , 
\end{equation}
where $i$ and $j$ are two data points and the exponential function maps the $distance$ to a range of [0, 1] with a lower $distance$ leading to a higher $Similarity$.
The $distance$ can be any kind of distance measurement between data points. In this study \add{we employed Euclidean distance for simplicity, familiarity, and continuity with prior work~\cite{kaul2021improving},} 
\begin{equation}
\label{eq:ecudis}
    distance(i,j) = \sqrt{(|x_i - x_j|)^2 + (|y_i - y_j|)^2}.
\end{equation}
\add{Note that the distance measure should be calculated based on all dimensions in the datasets for \autoref{eq:ecudis}.}

Although the results in this paper use the Euclidean distance, other distance measures may be more appropriate for certain domain-specific applications, datasets with unequal importance between variables, or datasets containing large amounts of noise.
For example, methods such as the Mahalanobis distance~\cite{flores2021mahalanobis} or propensity score matching~\cite{chen2022best} could be applied for treatment analyses and other healthcare applications.
We therefore provide several commonly used distance measures in the counterfactual library used to implement our guidance system~\cite{wang2024framework}. Developers can also implement or use their own preferred distance measures to replace the implemented ones if they have specific needs.

For typical analyses, users are seeking filter variables that indicate differences in outcomes between subsets. Therefore, the more similar CF is to IN, the less likely it is that the filter constraints is one of interest in the analysis, and vice versa.
We therefore define $D_{IN, CF}$, as introduced in \autoref{sec:definesubset}, as the normalized dissimilarity between IN and CF:
\begin{equation}
\label{eq:cfmeasure}
    D_{IN, CF} = 
    \frac{1}{|S_{IN}||S_{CF}|} 
    \sum_{i \in S_{IN}}^{|S_{IN}|} \sum_{j \in S_{CF}}^{|S_{CF}|}
    (1-Similarity(i,j)),
\end{equation}
where $S_{IN}$ and $S_{CF}$ are the IN and CF subsets, and $1-Similarity(i-j)$ is the dissimilarity between points $i$ and $j$. This formulation of $D_{IN, CF}$ results in a range of [0, 1].

\subsubsection{Remainder Dissimilarity}

The similarity between IN and REM also impacts data interpretation, as outcome differences between IN and REM could suggest the importance of non-filter variables in the dataset.
We therefore define the dissimilarity between IN and REM subsets $D_{IN, REM}$ in a similar way to \autoref{eq:cfmeasure} However, we replace the CF subset $S_{CF}$ with the REM subset $S_{REM}$ throughout the equation:
\begin{equation}
\label{eq:remmeasure}
    D_{IN, REM} = 
    \frac{1}{|S_{IN}||S_{REM}|} 
    \sum_{i \in S_{IN}}^{|S_{IN}|} \sum_{j \in S_{REM}}^{|S_{REM}|}
    (1-Similarity(i,j)),
\end{equation}

\subsubsection{Guidance Score}

To compute an overall guidance score, we incorporate $D_{IN, CF}$ and $D_{IN, REM}$ to represent how IN and CF are dissimilar and how IN and REM are dissimilar, respectively.
For each, larger values indicate that the selected filters may be of more importance to explore during guided analysis.

To reflect this design, we incorporate both dissimilarities together, weighting  $D_{IN, CF}$ more heavily than $D_{IN, REM}$ to capture the focus on the inclusion criteria, as:
\begin{equation}
\label{eq:relation}
    Guidance_{CF} = \frac{1}{2} (D_{IN, CF} + \sqrt{D_{IN, CF}*D_{IN, REM}}).
\end{equation}

To reduce the weight of  REM we calculate the geometric mean with the CF subset (i.e., the square root item).
This ensures that REM is impactful only when CF is impactful. This guides away from archetype case 2. 
With this guidance equation, we combine the impacts of both subsets, while emphasizing the impact of the CF subset.

As a baseline for our control group in our evaluation study, we also developed a correlation-based guidance measure. Correlation guidance ($Guidance_{corr}$) follows a somewhat similar computation process in our prototype, but replaces the calculation of counterfactual subset distances with correlations between IN and EX.

\subsubsection{Subset Distribution Score}

A threshold-based method is used to create data subsets.
The size $n$ of IN is directly determined by user-selected filters.
To create CF we select the $n$ closest point to IN from EX, resulting in IN and CF having the same size, with the remaining points belonging to REM, following previous work~\cite{wang2024empirical}.
However, if IN contains more than $\frac{1}{3}$ of all data points, we split the data points in EX evenly between CF and REM.

We note that the effectiveness of these subsets may be impacted by the relative sizes of their data samples.
For example, when IN includes almost all data samples from the target dataset, no matter how large the $Guidance_{CF}$ or $Guidance_{corr}$ value may be, we cannot conclude that the data subsets would have a high impact.
Similarly, if the size of IN were very small, the dissimilarity between these subsets would not be very informative.

Further, we define subset distribution scores $Distribution_{S1, S2}$ to measure the difference between the sizes of different subsets, based on the $SizeDifference_{S1, S2}$:
\begin{equation}
\label{eq:sizediff}
    SizeDifference_{S1, S2} = \frac{|S_{S2}|}{|S_{S2}|+|S_{S1}|},
\end{equation}

\begin{align}
\label{eq:distribution}
    Distribution_{S1, S2} = 1-2*|SizeDifference_{S1, S2} - \frac{1}{2}| \\
   \add{= 1-2*|\frac{|S_{S2}|-|S_{S1}|}{|S_{S2}|+|S_{S1}|}|},
\end{align}
which results in a normalized value between 0 and 1.

For example, $Guidance_{CF}$ 
relies primarily on the differences between IN and CF,
so we use $SizeDifference_{IN, CF}$.
In ideal cases, the number of samples in IN and CF would be similar, such that $SizeDifference_{IN, CF}$ would be close to 0.5, and the overall $Distribution_{IN, CF}$ would be close to 1.
Whenever IN has an extremely large or small number of samples, $SizeDifference_{IN, CF}$ will approach 0 or 1, and $Distribution_{IN, CF}$ will approach 0.
Therefore, this measure can be employed as an empirical validation of $Guidance_{IN, CF}$, where lower values of $Distribution_{IN, CF}$ imply a smaller impact of the CF subset.
Empirically, when $Distribution_{IN, CF}$ is smaller than 0.1, we find that the subset distribution cannot reliably support insights from $Guidance_{IN, CF}$.

For correlation-based guidance, since $Guidance_{corr}$ measures the differences between IN and EX,  we calculated its subset distribution score as $Distribution_{IN, EX}$.

\subsection{Implementation}
A Python implementation of the proposed counterfactual guidance technique can be found in the \emph{cf\_guidance} file accompanied by the \emph{Co-op} library~\cite{wang2024framework} as two functions, \emph{get\_cf\_guidance\_score} and \emph{get\_distribution\_score}. This library also contains basic computation mechanisms for creating counterfactual subsets, built on efficient scientific computing packages including NumPy, SciPy, and Pandas.
The integrated open-source library is available at \href{https://github.com/VACLab/Co-op}{\textcolor[RGB]{0,0,255}{GitHub}}.

\section{User Study}
\label{sec-evaluation}

We conducted a comparative user study using a prototype visual interface to evaluate the performance of counterfactual guidance and compare it to correlation-based guidance. The user study was approved by the UNC-Chapel Hill Institutional Review Board. This section provides detailed descriptions of the study design, analysis process, and results.

\subsection{Visual Interface}
\label{sec-prototype}

We designed the functionalities and interactions of the prototype system based on insights from a prior counterfactual-based exploration system~\cite{kaul2021improving} and expert interviews.
The system enables guided exploration by providing a feature guidance view to help users pick interesting variables and an analytical summary view to help them explore selected filters in detail.
Here we describe the two primary views, feature guidance and analytical details, and supported atomic interactions. We compare it with the \emph{CoFact}~\cite{kaul2021improving} interface to illustrate how our interface can provide a simplified counterfactual-guided exploration experience.

Note that following prior studies~\cite{yen2019exploratory, battle2019characterizing}, the task, dataset, and outcome variable were fixed in our study design (see \autoref{sec:data} and \autoref{sec:task}), therefore we disabled the configuration page for selecting datasets and outcome variables during the study. The same outcome variable is therefore always displayed, and the user is able to filter based on other variables in the dataset to examine changes in the distribution of this outcome variable and interpret relationships between these variables and the outcome variable.

\begin{figure*}[htbp]
\centering
\includegraphics[width=1.9\columnwidth]{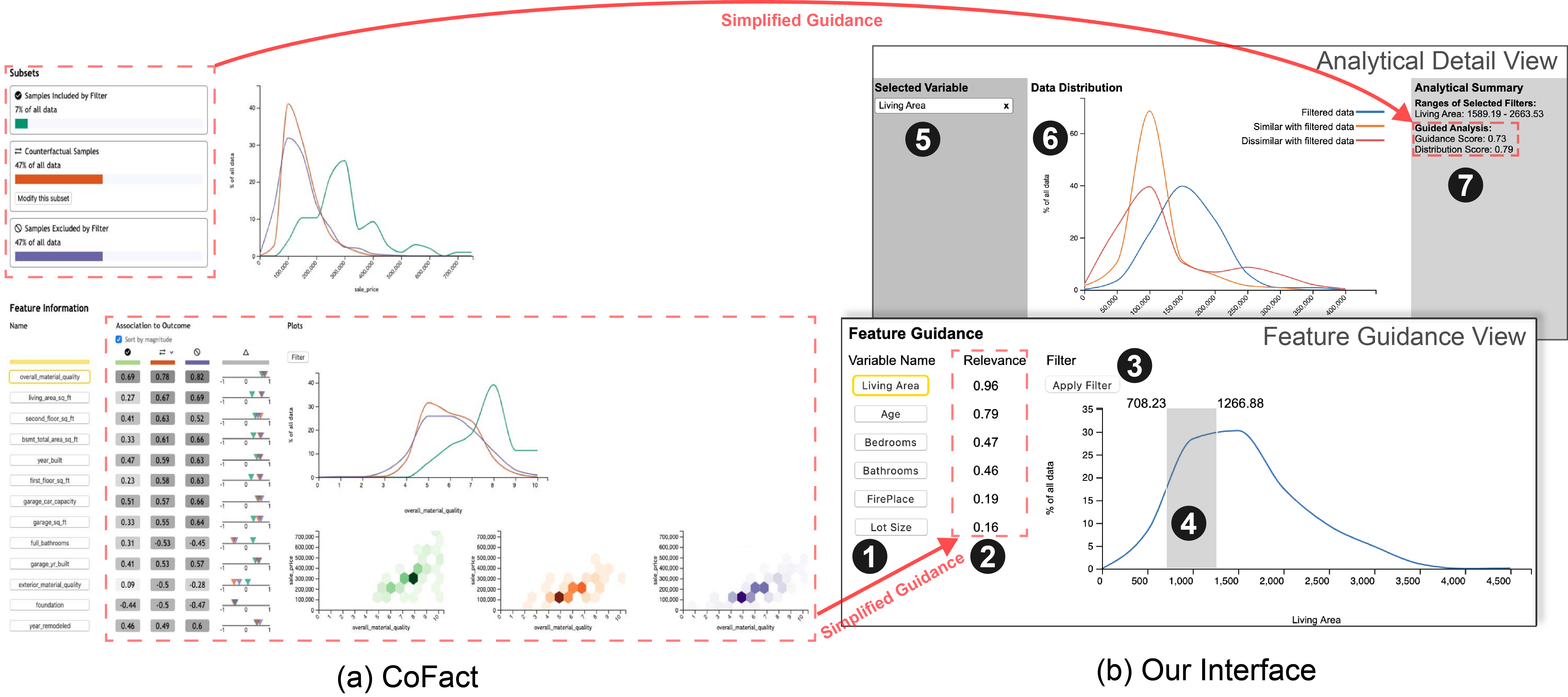}
\caption{ 
A comparison between \emph{CoFact}~\cite{kaul2021improving} (a) and our interface (b).
The \textcolor{red}{red} indicates the visualized counterfactual information shown to guide user analysis in each interface, demonstrating how our technique simplifies the counterfactual information shown to users.
The labels (b.1) to (b.7) refer to different information and functionality shown in the interface, see \autoref{sec-prototype} for details.
}
\label{fig:interface}
\vspace{-1em}
\end{figure*}

\paragraph{Feature Guidance.}
The bottom view in \autoref{fig:interface} (b) shows the feature guidance view for users to select variables for filtering.
Selectable variables (i.e., those other than the outcome and already selected variables) are shown under \emph{Variable Name}, see \autoref{fig:interface} (b.1).
The variables are ordered by their guidance values, shown in \emph{Relevance}, see \autoref{fig:interface} (b.2).
These values may reflect counterfactual guidance or correlation-based guidance, based on different user groups in the study.
The term \textit{Relevance} is used for both guidance types.
Users can select a variable name displayed in \autoref{fig:interface} (b.1), and then a visualization of the distribution of the selected variable will be shown in \emph{Filter}, see \autoref{fig:interface} (b.3).
In this \emph{Filter} view, users can click the distribution chart to control the filter ranges they want to apply, as shown in \autoref{fig:interface} (b.4), and click the \emph{Apply Filter} button to apply the selected filter range.
Once the button is clicked, the feature guidance view will be updated based on guidance values calculated using user-applied filters.

\paragraph{Analytical Detail.}
After applying filters in the feature guidance view, users can switch to a more detailed view, as shown at the top of \autoref{fig:interface} (b).
Selected variables are shown in the \emph{Selected Variable} panel, see \autoref{fig:interface} (b.5), in the order they were added.
Users can also remove a selected variable by clicking the $\times$ button next to each variable name in \autoref{fig:interface} (b.5).
Data distributions of the outcome variable for IN, EX, and CF subsets are shown in the \emph{Data Distribution} panel, see \autoref{fig:interface} (b.6).
Note that to ease understanding of the subset distributions, we did not show the subset names explicitly, instead explaining the IN, CF, and REM subsets' patterns as \emph{filtered data}, \emph{those similar with filtered data}, and \emph{those dissimilar with filtered data}, as shown in the data legend in \autoref{fig:interface} (b.6).
Similarly, when using correlation-based guidance, the distributions of IN and EX are shown, where the legend of EX shows \emph{those not in filtered data}.
Detailed analytical and guidance information, including all filter ranges and guidance values (guidance score and subset distribution score), are shown in the \emph{Analytical Summary} panel \autoref{fig:interface} (b.7).

\paragraph{Atomic Interaction Types.}
Two atomic interactions are available in the system.
\textbf{Changing filter variables} refers to user interaction to add or remove different variables from the filter inclusion criteria to explore their impact on the outcome.
\textbf{Changing filter ranges} refers to user interaction to adjust the range of values used for a filter variable.

\paragraph{Comparison with \emph{CoFact}~\cite{kaul2021improving}.}
\autoref{fig:interface} compares our interface with \emph{CoFact}.
The red dashed boxes emphasize the panels and views that show counterfactual and guidance information to help users explore data in each interface.
Our interface simplifies much of the information related to guidance, whereas 
\emph{CoFact} (see \autoref{fig:interface} (a)) shows more complex subset information and visualizations for achieving the same goal.

\subsection{Synthetic Data}
\label{sec:data}

One key limitation of prior empirical studies on counterfactuals in visualization is the lack of ground truth causal relations in the studied datasets.
To address this issue we generated a dataset with a defined ground truth causality between variables based on a causal graph by selecting example variable names from typical healthcare data~\cite{gotz2016data}, such as \emph{blood pressure} and \emph{cholesterol}.

Our data generation is based on prior work on graphical causal inference mechanisms.
Since constructing causal models from a dataset is usually complex and may not always guarantee ground-truth causality, we instead first defined a directed acyclic causal graph using our selected variables as nodes and assigned causal relationship strengths between these variables as links using \emph{DoWhy}~\cite{dowhy}.
We then generated the synthetic data based on the constructed causal graph~\cite{jarry2021quantitative} (see \href{https://github.com/VACLab/CausalSynth}{\textcolor[RGB]{0,0,255}{\emph{CausalSynth}}}~\cite{wang2024causalsynth} for a web application). Due to the difficulties in controlling the causal strength of categorical variables in a causal graph, all of the variables are continuous.

\begin{figure}[htbp]
\centering
\includegraphics[width=0.9\columnwidth]{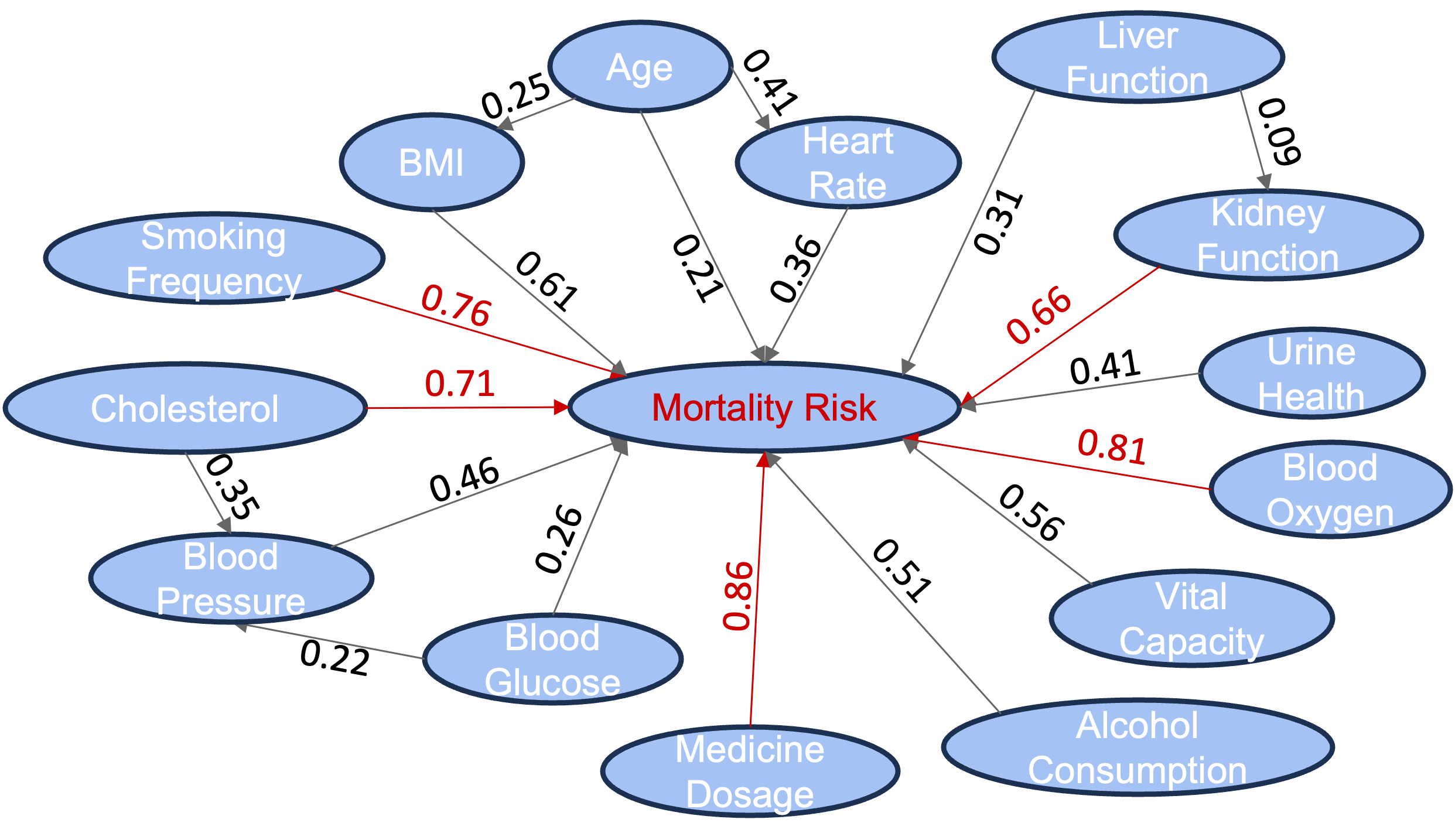}
\caption{ 
The defined casual graph in the synthetic data.
The middle node is the target outcome, \emph{mortality risk}, shown in red.
The values near each link are the causal relationship strengths.
The top 5 causal links are shown in red.
Causal strengths range from 0.21 to 0.86, with a 0.05 interval between each causal strength in the graph.
}
\label{fig:data}
\vspace{-1.5em}
\end{figure}

\autoref{fig:data} shows the defined causal graph and corresponding values of the ground truth causal relationships shown near each causal link.
In this causal graph, we define \emph{mortality risk} as the target outcome variable with different causal relationship strengths to other variables. The top five causal links are shown in red.
The causal relationship strengths are assigned with a 0.05 interval (i.e., one's strength to outcome is 0.05 higher than its nearest-lower factor) to avoid potential effects caused by unbalanced differences in causal strength.
Several causal links between factors were randomly added to increase data complexity (e.g., \emph{cholesterol} $\rightarrow$ \emph{blood pressure}).

To facilitate exploratory analysis, we predefined default filter ranges for each variable, determined by two expert analysts to highlight potential interesting filter ranges to serve as inclusion criteria for IN.

\subsection{Participants}

We recruited 20 users (14 male, 6 female; 19--30 years old) to participate in the study via mailing lists and contacts within professional networks.
All participants were at least 18 years old, had or were pursuing a university degree, and had experience using visualization and data analysis such as taking information visualization or data science courses.
We employed a between-subjects design by randomly assigning 10 participants to use counterfactual guidance (the CFACT group) while the other 10 participants used correlation guidance (the CORR group) in the same prototype system.
On average, the study took around 20-30 minutes for users in each group.

\subsection{Hypotheses}

We aimed to explore the effectiveness of counterfactual guidance for finding causal relationships in a dataset using exploratory visual analysis.
Based on this goal, we hypothesized that:

\begin{itemize}
    \item \textbf{H1:} Counterfactual guidance would lead to higher accuracy in finding causal relationships compared to correlation guidance.
    \item \textbf{H2:} Counterfactual guidance would lead to higher confidence in users' findings.
    \item \textbf{H3:} Counterfactual guidance would lead to fewer wrong attempts.
    \item \textbf{H4:} Users using counterfactual guidance may spend more time in analysis.
\end{itemize}

Through these hypotheses, our study aims to better characterize the impact of counterfactual guidance and compare its effectiveness with correlation-based guidance.

\subsection{Procedure and Task}
\label{sec:task}

After accepting the informed consent form including the study's purpose and participants' rights, we gave the users a tour of the visual interface and introduced the available interactions and functionalities.
With \emph{mortality risk} chosen as the target outcome variable for the visual interface, participants were asked to complete two tasks:
\begin{itemize}
    \item \textbf{T1:} Identify which variables may be most likely to cause higher \emph{mortality risk}, choosing up to 5 variables.
    \item \textbf{T2:} Rank those variables from most likely to least likely.
\end{itemize}
Participants recorded their confidence on a 5-point Likert scale following each task.
The ground truth results of these tasks are five variables with the highest causal relationship strengths to \emph{mortality risk}, as shown by the red links in \autoref{fig:data}.

\subsection{Analysis}
We measured our results as accuracy, confidence, interactions, and time spent.
The only independent factor was the two user groups, so we calculated individual t-tests based on these measures.

\subsection{Results}
\label{sec-study-res}
\autoref{tab:test} provides the results of the main measures from our analysis.
Of note is that all users reported 5 variables, despite the freedom to choose fewer. 

\begin{table}[t]
\centering
\caption{The t-tests results for main measures in our study. Significant effects are indicated by bold text and the corresponding rows are highlighted in green. }
\vspace{-0.5em}
\includegraphics[width=0.8\linewidth]{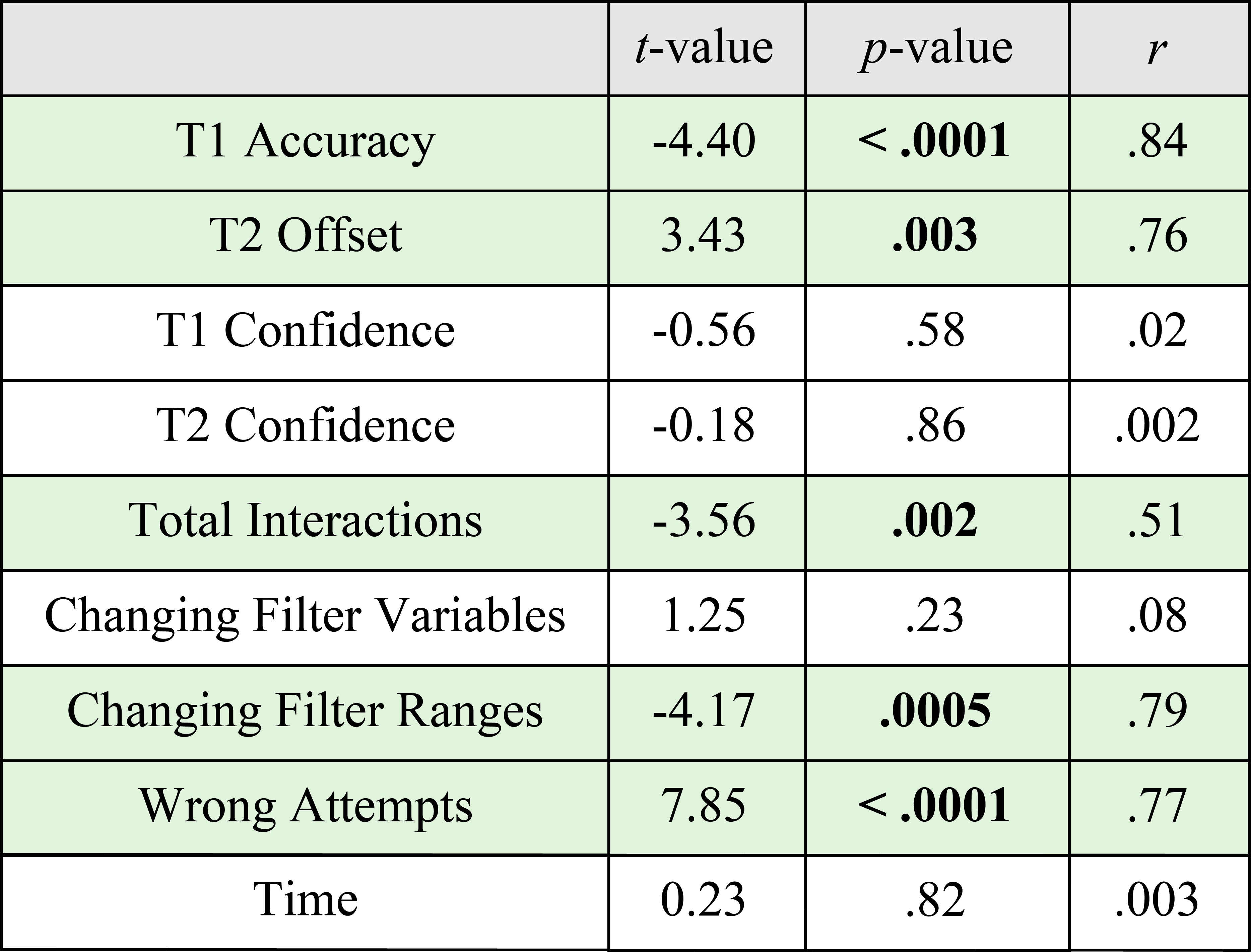} 
\label{tab:test}
\vspace{-1.5em}
\end{table}

\subsubsection{Accuracy}
We assess the accuracy of \textbf{T1} by computing the relative ratio between the number of correct answers and 5, i.e., the number of provided answers that were in the top 5 of the ground truth causal relationships, independent of order.

To measure the accuracy for \textbf{T2} we use an offset distance, similar to the edit distance~\cite{ristad1998learning}:
\begin{equation}
    Offset_{T2} = \sum_{i=1}^{5} |i - Rank_{GT}(Answer_i)|,
\end{equation}
where $Answer_i$ is the variable at the $i$-th position ($i \in [1,5]$) in the user's ranking, and $Rank_{GT}(Answer_i)$ is the ground truth ($GT$) ranking of this variable.
This metric measures the accuracy between two rankings by summing how much each variable's position in the user ranking deviates from its correct position in the ground truth ranking.

We found significant differences in accuracy between the CFACT and CORR groups for both \textbf{T1} ($t = -4.40, p < .0001$) and \textbf{T2} ($t = 3.43, p = .003$).
The mean differences between CFACT and CORR are 0.28 for \textbf{T1} and 2.80 for \textbf{T2}, indicating that counterfactual guidance performed better. The results on accuracy therefore support \textbf{H1}: we found that counterfactual guidance can lead to higher accuracy in finding causal relationships.

\subsubsection{Confidence}
We did not find any significant differences for confidence between the CFACT and CORR groups.
This indicates that the results do not support \textbf{H2}: we found users' confidence was not significantly impacted by guidance type.

\subsubsection{Atomic Interactions}
\label{sec:res_atomic}
We also examined differences with respect to the overall atomic interactions (changing filter variables and changing filter ranges) employed by users in each group. Users in CFACT employed significantly more atomic interactions ($t = -3.56, p = .002$). We also found that the CFACT group performed significantly more changes to filter ranges ($t = -4.17, p = .0005$), but found no significant difference in changing filter variables.

Furthermore, we identified wrong attempts as the total number of atomic interactions involving a variable that does not exist in the top 5 ground truth causal variables.
We found a significant difference in wrong attempts ($t = 7.85, p = <.0001$), with a mean difference of -8.7, indicating that users in CFACT had fewer wrong attempts.
Our results on interactions therefore support \textbf{H3}: we found that counterfactual guidance can help users explore fewer incorrect variables.

\subsubsection{Time}
We found no significant difference between groups in the overall time spent on analysis, indicating that the results do not support \textbf{H4}: counterfactual guidance did not lead to a significant difference in time spent.

\begin{table}[t]
\centering
\caption{Significance results for different interaction behaviors in our study.  Significant effects are indicated by bold text and the corresponding rows are highlighted in green.}
\vspace{-0.5em}
\includegraphics[width=0.8\linewidth]{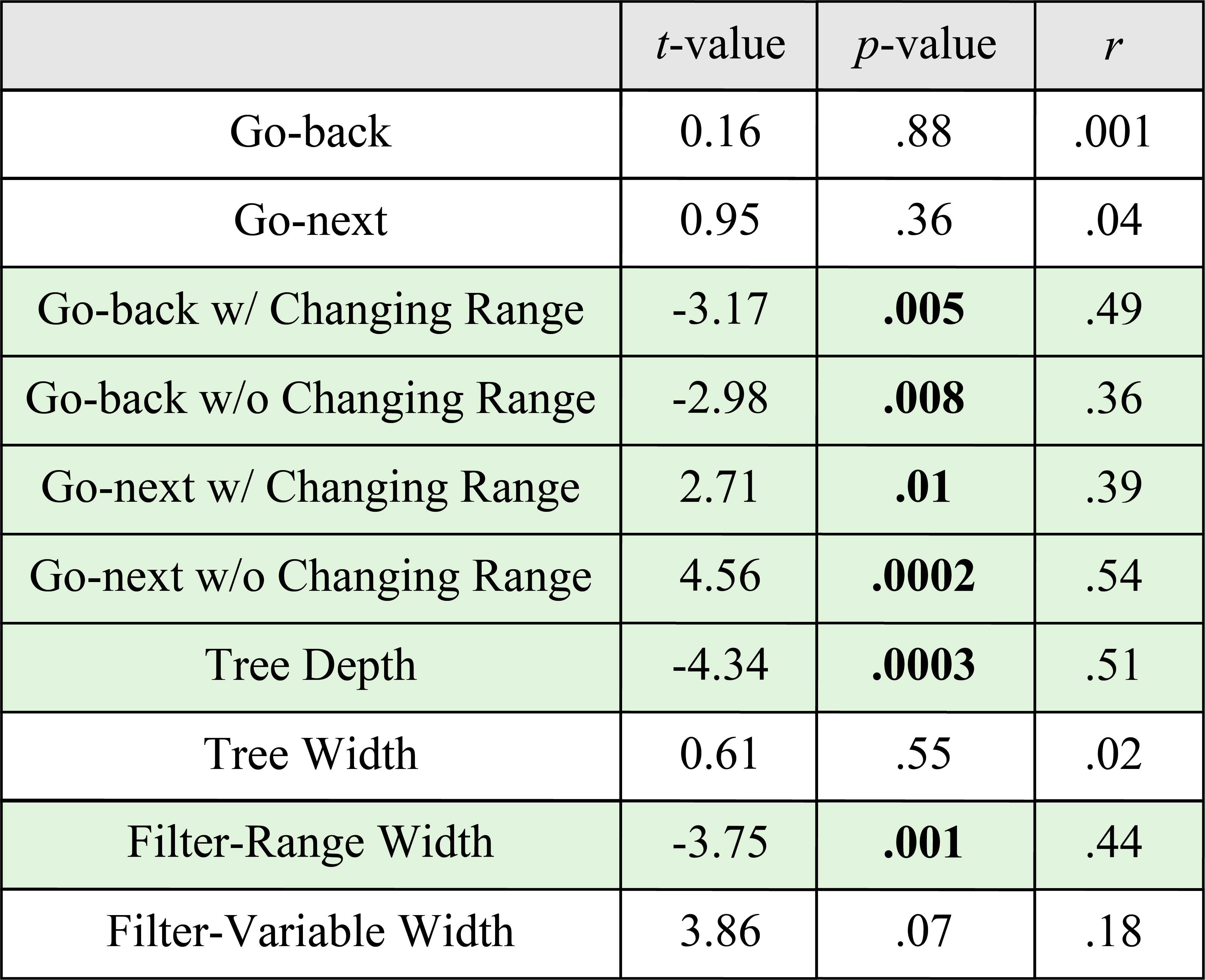} 
\label{tab:behavior}
\vspace{-1.5em}
\end{table}

\subsubsection{Exploratory Analysis for Interaction Behaviors}

Interaction strategies and behaviors can also impact users' exploration~\cite{battle2019characterizing, yen2019exploratory}. 
Given the significant differences in accuracy and number of interactions, we performed a more fine-grained analysis of detailed interaction behaviors to identify any differences between the CFACT and CORR groups.
This section follows the previous definition of atomic interaction types from \autoref{sec-prototype} and extends them based on user behaviors.
See \autoref{tab:behavior} for a summary of the results for each interaction behavior.

\begin{figure}[b]
\centering
\vspace{-1em}
\includegraphics[width=0.8\columnwidth]{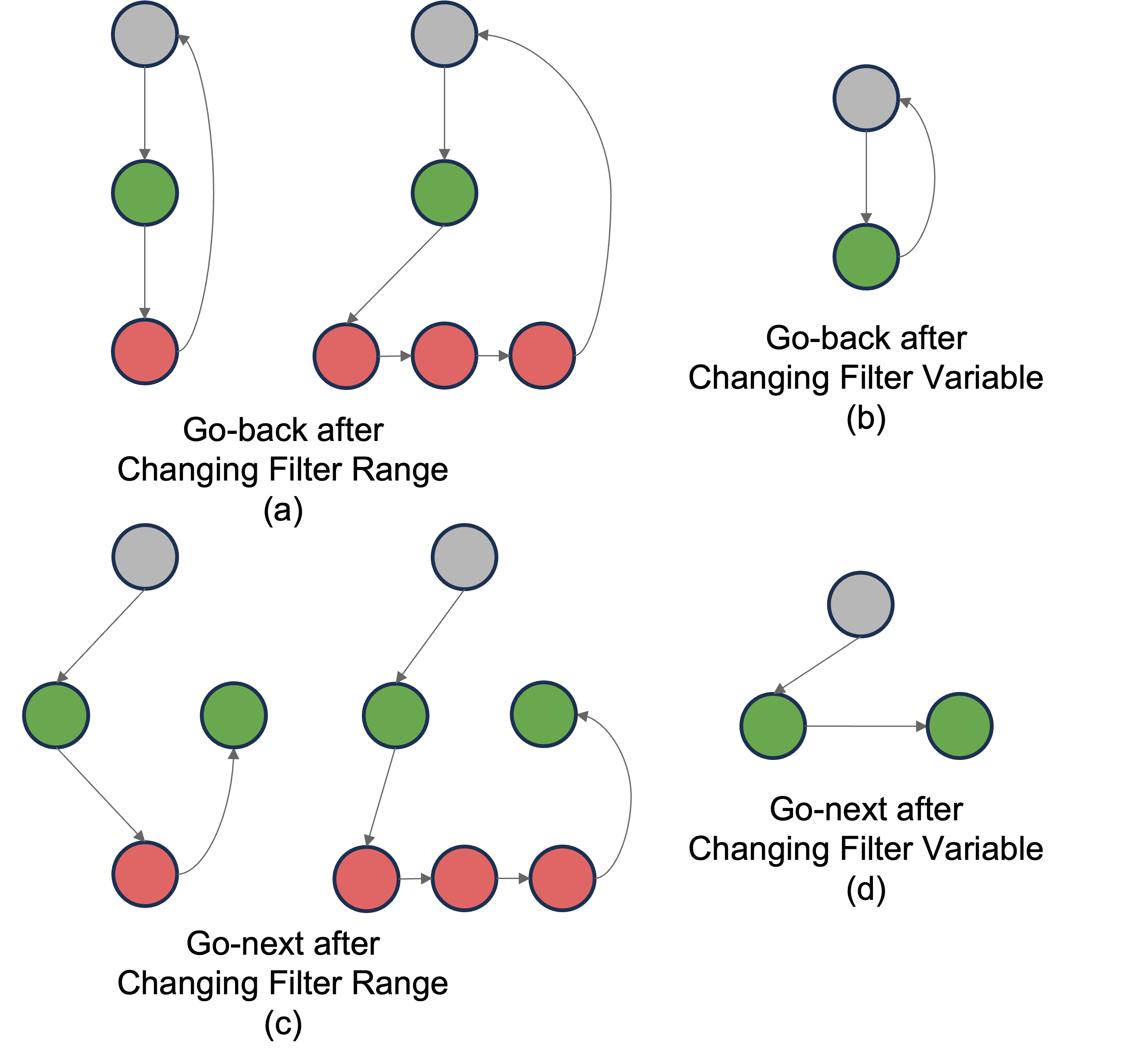}
\vspace{-0.5em}
\caption{ 
Four types of identified interaction behaviors in our exploratory analysis consist of atomic interactions. (a-b) are go-back behaviors and (c-d) are go-next behaviors.
}
\vspace{-1.5em}
\label{fig:interaction}
\end{figure}

\autoref{fig:interaction} illustrates two sets of key interaction behaviors---go-back and go-next---which we identified from users' exploration behaviors.
Each node represents an atomic interaction (changing filter variables or filter ranges).
The gray nodes are starting points, which could be any interaction.
The green nodes represent a filter variable interaction and the red nodes represent  a filter range interaction.
Go-back behaviors are cases in which users first add a variable, and then remove it after exploring insights with this variable.
Go-next behaviors are cases in which users add a second variable during the analysis of the current variable.
Go-next may therefore indicate that users are trying to explore the combined impact of multiple variables.
In addition, we subdivide go-back and go-next interaction behaviors into two subtypes according to users' filter range interactions.
Go-back interactions include go-back after changing filter ranges (\autoref{fig:interaction} (a)) and go-back without changing filter ranges (\autoref{fig:interaction} (b)). 
The same subtypes are also included for go-next (\autoref{fig:interaction} (c, d)).

No significant differences were found between the overall go-back and go-next behaviors.
However, significant differences were found between each sub-type.
The CFACT group exhibited more interaction behaviors of go-back ($t=-3.17,p = .005$) and go-next  ($t=2.71,p = .01$) after changing filter ranges, with mean differences of 1.8 and 1.2 respectively, and exhibited fewer interaction behaviors of go-back ($t=-2.98,p = .008$) and go-next ($t=4.56,p = .0002$) after changing filter variables, with mean differences of -1.8 and -2.2 respectively.

\begin{figure}[t]
\centering
\includegraphics[width=0.65\columnwidth]{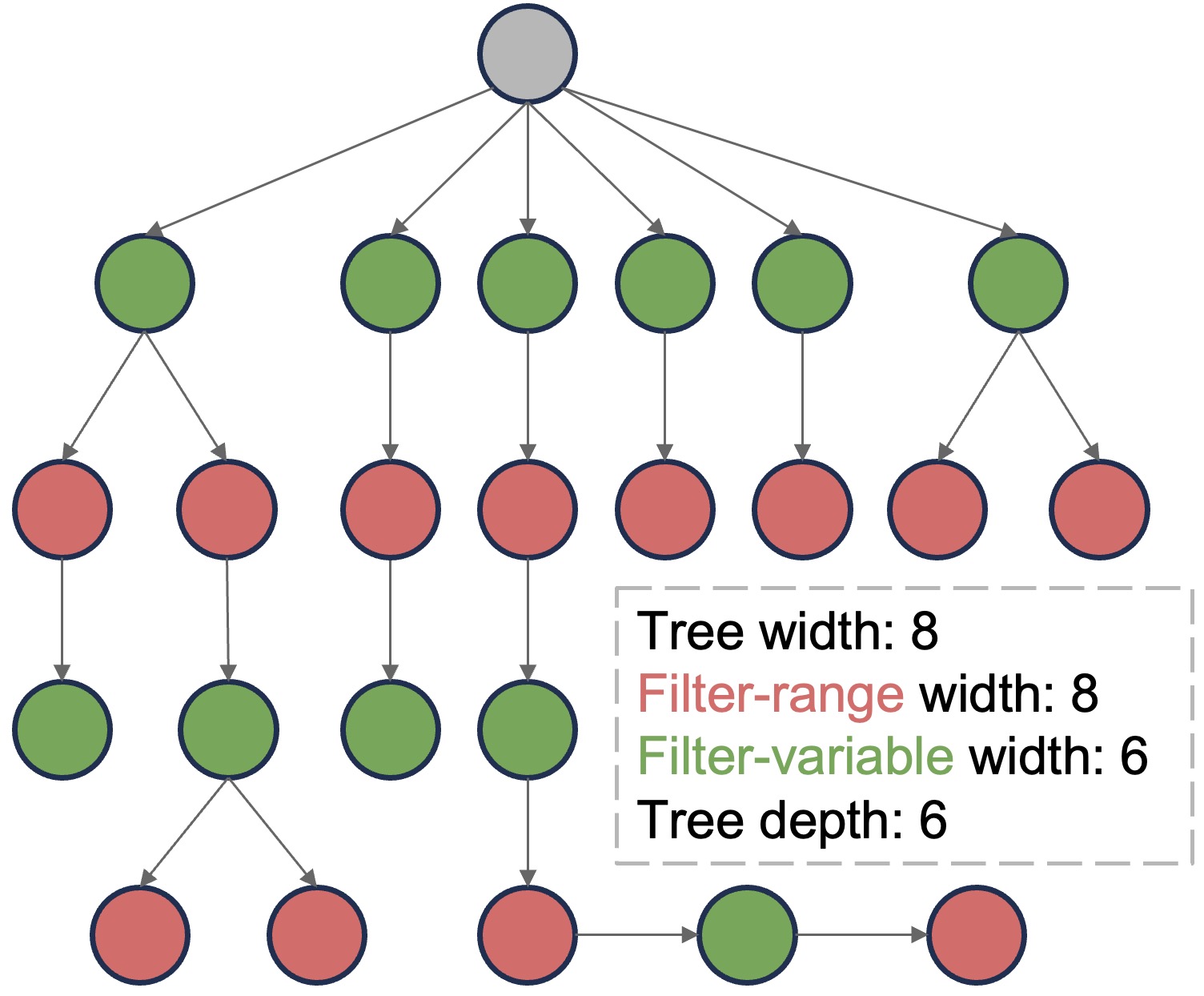}
\caption{ 
A search tree of a user in the CFACT group, with a width of 8, a filter-range layer of 8, a filter-variable layer width of 6, and a depth of 6.
}
\label{fig:tree}
\vspace{-1.5em}
\end{figure}

Additionally, we examined the impact of counterfactual guidance on the overall interaction structures of users.
Following prior work~\cite{battle2019characterizing}, we identified each variable change and range change as a node and removed the backward edges to construct search trees for each user and assessed the data patterns.
The tree structure captures the exploration path and helps us better evaluate users' analysis behaviors~\cite{battle2019characterizing}.
\autoref{fig:tree} illustrates a search tree created based on a user in the CFACT group of our study, with red and green nodes denoting changing filter variables and changing filter ranges respectively.
We denote the layer with red nodes as the filter-range layer and the layer with green nodes as the filter-variable layer.

We employed tree width and height as measurements~\cite{battle2019characterizing} to perform significance tests between the CFACT and CORR groups, using the largest width and the largest height, as well as the largest width on the filter-range layer (filter-range width) and the filter-variable layer (filter-variable width), for each tree (see \autoref{fig:tree}).
We found a significant difference for tree depth ($t=-4.34, p = .0003$), with a 2.5 mean difference, indicating that users in CFACT had deeper search trees.
For tree width we found a significant difference for the filter-range layer ($t=-3.75, p = .001$), with a 2.2 mean difference, indicating that users in CFACT had wider filter-range layer widths. 

Taken together, these exploratory patterns indicate that counterfactual guidance could lead to more filter range explorations within each filter variable, and visualizations of more filter variable combinations.

\section{Discussion}
\label{sec-discussion}

The analysis results indicate that counterfactual guidance was effective for improving the accuracy of finding causal relationships in exploratory analysis, and that counterfactual guidance led to different interaction behaviors compared to correlation-based guidance.
Here we discuss how our results relate to prior work,  
identify limitations of this work, and point out future research directions.

\subsection{Reflection on Prior Work}
Our results indicate the effectiveness of using counterfactuals as guidance in exploratory visual analysis workflows.
Here we discuss how our results reflect on insights from prior work.

\subsubsection{Counterfactual Visualization}
The main insights of general-purpose counterfactual visualizations were reported by Kaul et al.~\cite{kaul2021improving} and Wang et al.~\cite{wang2024empirical}, which both demonstrate the potential utility of counterfactuals in fostering a deeper understanding of causal relationships within datasets.
Our method, even though largely simplifying the counterfactual information in the interface, also found that using counterfactual information as guidance can benefit data exploration.
One of the reported limitations of these studies was increased analysis time and complexity. Our counterfactual guidance method was able to address these limitations while retaining the benefits of counterfactuals for analyzing causal relationships.

However, our results are inconsistent with their findings in terms of confidence.
Kaul et al.~\cite{kaul2021improving} reported counterfactual visualizations would reduce or confirm users' confidence in different filter choices in analytics systems and Wang et al.~\cite{wang2024empirical} also found counterfactuals would impact confidence in static charts.
We anticipate that these contradictions may be a result of our simplification of counterfactual information, as shown in \autoref{fig:interface}.
But further work should better examine this assumption.

\subsubsection{Exploratory Behaviors}

Our results indicate that when exploring data with counterfactual guidance, users' interactions are more likely to construct deeper search trees.
This insight aligns with Battle and Heer~\cite{battle2019characterizing}, where they found exploratory sessions are more likely to be depth-oriented (using Tableau).

Moreover, we also found correlation-based guidance leads to a relatively low overall causal inference accuracy.
This result aligns with Zgraggen et al.~\cite{zgraggen2018investigating}, however, Battle and Heer~\cite{battle2019characterizing} found that users performed very well in their tasks.
They explain these differences as being due to recruiting more experienced users for their datasets (i.e., from professional Tableau User Groups).
In addition, we propose that these differences may also be due to differences in tasks and datasets, for instance, our task is more open-ended, similar to \cite{zgraggen2018investigating}.

\subsection{Limitations and Future Directions}

This research, while providing valuable insights into the use of counterfactual guidance in visual analytics, acknowledges several limitations that pave the way for future exploration:

\subsubsection{Empirical Elements in Counterfactual Guidance Computation}
When computing counterfactual guidance, we mitigated the REM subset's impact using the geometric mean ($\sqrt{D_{IN, CF}*D_{IN, REM}}$), and added it to the impact of the CF subset ($D_{IN, CF}$) in our counterfactual guidance technique. This formulation was determined empirically.
Future work should aim to provide analytical evidence to support these weight choices or explore adaptive weighting schemes to better emphasize the significance of the impacts of the two subsets.
In addition, \add{for the subset size and subset distribution score}, we provided empirical suggestions to users, but did not perform any statistical analysis to \add{measure at which scale the subset size and its distribution score may imply a significant impact for the effectiveness of the guidance.}
More statistical analysis and usage suggestions \add{for empirical thresholds related to the data subsets should be thoroughly examined} in future work.

\subsubsection{Automated Computing Counterfactual Filtering}
The counterfactual guidance technique does not incorporate automated methods for determining filter ranges for computing counterfactual subsets, therefore we employed expert-designed filters as the default for both counterfactual and correlation guidance.
This limits the potential usage of guidance as an overview of the dataset prior to any user interactions.
Future improvements should integrate automated approaches to suggest default filter ranges, such as machine learning models, which may enhance the guidance’s ability to create meaningful counterfactual subsets automatically, leading to more efficient data exploration.

\subsubsection{Synthetic Data}
The synthetic data used cannot accurately reflect the complex causal relationships as well as noise existing in real-world datasets~\cite{GentzelGJ19}.
\add{The noise in these datasets may make counterfactual analysis more challenging.}
Future studies should investigate the applicability of our guidance technique to more real-world and complex datasets.
\add{Further, other distance measures such as Mahalanobis distance~\cite{de2000mahalanobis} should be explored in the future to mitigate the impact of noise.}
\add{Further, real-world data is often influenced by numerous factors beyond immediate causal relations between two variables.
These factors, such as confounders and colliders, can introduce significant biases and distortions into users' analyses, potentially leading to incorrect conclusions~\cite{greenland2003quantifying}.}
However, the synthetic data did not account for confounders and colliders, limiting its ability to simulate real-world cases.
Addressing these elements is essential for a more comprehensive evaluation of the counterfactual guidance method.
\add{In the future, we plan to incorporate more causal structure techniques to better identify confounders and colliders, such as causal diagrams, statistical testing, and domain knowledge-assisted identification~\cite{sjolander2017confounders}.}
Furthermore, our study did not test unbalanced distributions or out-of-distribution cases between the employed subsets.
Investigating these scenarios is important for understanding the robustness of counterfactual guidance.

\subsubsection{Task Design}
Our study tested only open-ended tasks even though it required users to explore many variables.
Providing more tasks and designing them with a more focused scope (e.g., ``\emph{What relationships do you observe involving weather conditions and strike frequency?}~\cite{battle2019characterizing}'') may reveal different exploratory insights.
\add{At the same time, experimenting with questions that are even more 
open-ended, such as inferential tasks, may better examine users' abilities to reason from data~\cite{10.2312:evs.20221086}.}
Future research should explore a wider range of exploration tasks \add{designed to be both more focused and more open-ended}.

\subsubsection{Impact of Prior Beliefs}
Our study did not consider the impact of users’ pre-existing beliefs when selecting variables in the synthetic data, which may impact users' data interpretation.
For example, studies have shown that users' prior beliefs can impact their estimations of correlations~\cite{xiong2022seeing} and judgments of causalities~\cite{wang2024causal} from visualizations. Meanwhile, techniques such as Bayesian-guided inference can be effectively used to assist users' belief updating~\cite{kim2020bayesian}.
Future research should examine how beliefs influence users' exploration process when using counterfactual-based guidance, and investigate potential technologies to assist in evaluating and estimating priors.

\subsubsection{Visual Interface}
The employed exploratory visual interface did not support sophisticated visualizations or more complex exploratory interactions, \add{such as users’ random exploration strategies~\cite{battle2019characterizing}, varying data aggregation levels~\cite{xiong2019illusion}, and different visual encodings~\cite{quadri2024do}.} However, these factors may impact users' data understanding and exploratory behaviors.
Enhancing the interface to accommodate a wider range of interactions, user customizations, and visualizations could improve user experience and analytical outcomes and may lead to more interesting interaction behavior patterns, which should be explored in future work.

\section{Conclusion}
\label{sec-conclusion}

This paper presented a novel counterfactual-based guidance technique to support exploratory data analytics.
By transforming the subsets generated for counterfactual visualizations into guidance values, our approach addresses the limitations of complexity and time-consuming analysis associated with counterfactual visualizations.
Employing counterfactual guidance, we retain the advantages of counterfactual reasoning while mitigating the cognitive load on users.
Through the usage scenario analysis and empirical study, our investigation has demonstrated the efficacy of counterfactual guidance for exploratory visual analysis.
The empirical evidence shows that counterfactual guidance significantly enhances the accuracy of interpreting causal relationships in datasets, suggesting that counterfactual guidance can serve as a powerful tool in guided visual analytics without overwhelming users.
In summary, our research advocates for the integration of counterfactual reasoning into visual analytics systems more widely using counterfactual guidance.
This technique promises to facilitate more precise and insightful data exploration, ultimately empowering users to make informed decisions based on robust exploratory analysis and causal inferences.

\acknowledgments{
We thank the reviewers for their insightful comments.
This material is based upon work supported by the National Science Foundation under Grant No. 2211845.
}

\bibliographystyle{abbrv-doi-hyperref}

\bibliography{main}
\end{document}